\documentclass[aip, reprint]{revtex4-2}
\usepackage{amssymb, amsthm, amsmath, amsfonts}
\usepackage{bm}
\usepackage[pdftex,dvipsnames,usenames]{xcolor}
\usepackage{graphicx, mathptmx}
\usepackage[colorlinks=true,urlcolor=blue,citecolor=blue,linkcolor=blue]{hyperref}
\usepackage{subcaption}
\usepackage[normalem]{ulem}
\usepackage{upgreek}
\usepackage{booktabs}
\usepackage{colortbl}
\usepackage{multirow}
\usepackage{enumitem}
\newcommand{\hlem}[1]{\textcolor{DarkOrchid}{#1}}
\DeclareMathAlphabet{\mathcal}{OMS}{cmsy}{m}{n}
\graphicspath{ {./Images/} }
\usepackage{blindtext}

\begin{document}

\title{Optimized diamond inverted nanocones for enhanced color center to fiber coupling}
\author{Cem G\"uney Torun}
\affiliation{Integrated Quantum Photonics, Institute for Physics, Humboldt-Universit\"at zu Berlin, Newtonstraße 15, 12489 Berlin, Germany}
\author{Philipp-Immanuel Schneider}
\affiliation{JCMwave GmbH, Bolivarallee 22, D-14050 Berlin, Germany}
\affiliation{Zuse Institute Berlin (ZIB), Takustraße 7, D-14195 Berlin, Germany}
\author{Martin Hammerschmidt}
\affiliation{JCMwave GmbH, Bolivarallee 22, D-14050 Berlin, Germany}
\affiliation{Zuse Institute Berlin (ZIB), Takustraße 7, D-14195 Berlin, Germany}
\author{Sven Burger}
\affiliation{JCMwave GmbH, Bolivarallee 22, D-14050 Berlin, Germany}
\affiliation{Zuse Institute Berlin (ZIB), Takustraße 7, D-14195 Berlin, Germany}
\author{Joseph H.D. Munns}
\affiliation{Integrated Quantum Photonics, Institute for Physics, Humboldt-Universit\"at zu Berlin, Newtonstraße 15, 12489 Berlin, Germany}
\author{Tim Schr\"oder}
\email{tim.schroeder@physik.hu-berlin.de}
\affiliation{Integrated Quantum Photonics, Institute for Physics, Humboldt-Universit\"at zu Berlin, Newtonstraße 15, 12489 Berlin, Germany}
\affiliation{Diamond Nanophotonics, Ferdinand-Braun-Institut, Gustav-Kirchhoff-Straße 4, 12489 Berlin, Germany}
\date{\today}

\begin{abstract}
  Nanostructures can be used for boosting the light outcoupling of color centers in diamond; however, the fiber coupling performance of these nanostructures is rarely investigated.
    Here, we use a finite element method for computing the emission from color centers in inverted nanocones and the overlap of this emission with the propagation mode in a single-mode fiber.
    Using different figures of merit, the inverted nanocone parameters are optimized to obtain maximal fiber coupling efficiency, free-space collection efficiency, or rate enhancement.
    The optimized inverted nanocone designs show promising results with 66\% fiber coupling or 83\% free-space coupling efficiency at the tin-vacancy center zero-phonon line wavelength of 619 nm.
    Moreover, when evaluated for broadband performance, the optimized designs show 55\% and 76\% for fiber coupling and free-space efficiencies respectively, for collecting the full tin-vacancy emission spectrum at room temperature.
    An analysis of fabrication insensitivity indicates that these nanostructures are robust against imperfections.
    For maximum emission rate into a fiber mode, a design with a Purcell factor of 2.34 is identified.
    Finally, possible improvements offered by a hybrid inverted nanocone, formed by patterning into two different materials, are investigated, and increases the achievable fiber coupling efficiency to 71\%. 
\end{abstract}
\maketitle

\begin{figure}[t]
\includegraphics[width=\linewidth]{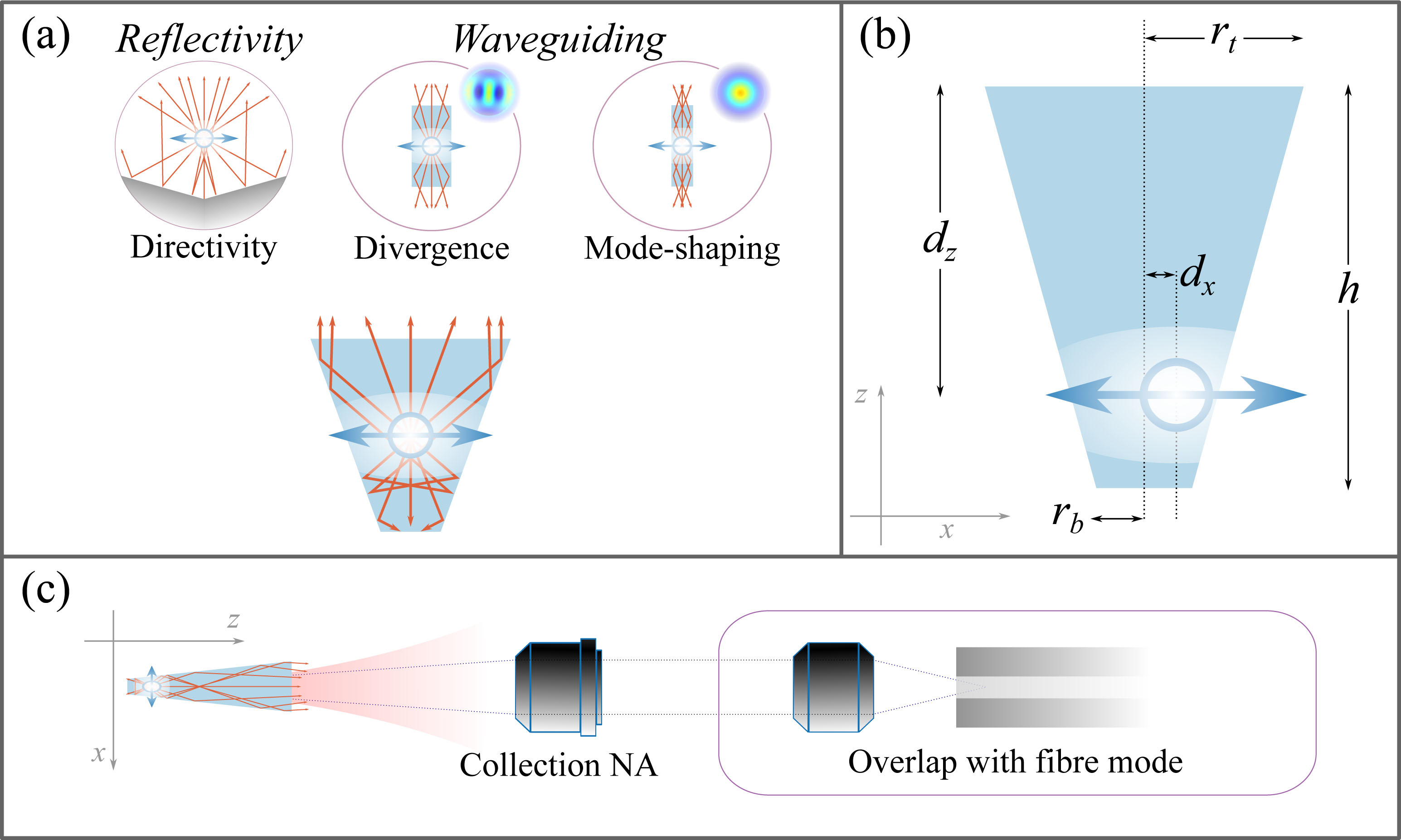}
\caption{
(a) Qualitative properties of nanostructures relevant for photon collection seen from analogy with a plane mirror, and nanopillar structures:
Directivity of the emission to the collection (upper) side; divergence determining emission into a particular solid angle; and mode shape determines how much of the emission can be collected into the fundamental mode of the waveguide.
(b) Dimensions used to parametrize INC design.
(c) Optical collection schemes for determining a figure of merit.
The free space collection efficiency ($\eta_\mathrm{FS}$) is defined as the fraction of emission that can be collected by the first collection optics.
The fiber coupling efficiency ($\eta_\mathrm{Fib}$) is defined as the fraction of emission coupling to the fiber after a second lens is used to image the emission onto the fiber facet.
}
\label{fig:INCdesign}
\end{figure}

Color centers in diamond are solid-state atom-like systems that are a promising platforms for quantum information processing \cite{Yao.2012, Childress.2013,Atature.2018} and quantum sensing \cite{Degen.2017,Wu.2016} due to their ability to encode optically addressable qubits.
However, challenges regarding the collection of emission from these centers remain a limiting factor.
The refractive index mismatch between diamond ($n=2.41$) and air ($n=1$) presents a critical angle of $24.5^\circ$ for total internal reflection.
This results in only around 4\% of the emission to outcouple from the sample to propagating air modes from an emitter in bulk diamond\cite{Hausmann.2010}.
Additionally, the dipole emission pattern of color centers is distributed into a solid angle of 4$\pi$ and thus, constitutes a mismatch with the targeted single fiber modes.

In order to overcome these limitations, fabricating optical nanostructures into diamond is a common approach\cite{Schroeder.2016}.
A wide range of strategies have been demonstrated, including applications with microlenses\cite{Hadden.2010,Jamali.2014}, nanopillars\cite{Babinec.2010,Hausmann.2010}, bullseye gratings\cite{Li.2015,Zheng.2017}, nanocones\cite{Momenzadeh.2015,Zhu.2020}, parabolic reflectors\cite{Wan.2018}, and recently, inverted nanocones\cite{Torun.2020,Jeon.2020}.
In order to quantify the efficiency achieved by these nanostructures, a {\em free-space collection efficiency} ($\eta_\mathrm{FS}$) figure of merit is generally employed.
This value is determined by the emission into a solid angle that is determined by the numerical aperture of the collection optics.
However, for quantum technology applications, coupling this emission to a single fiber mode is desirable such that multiple devices can be connected efficiently, and scalably.
Furthermore, for many quantum information processing algorithms, indistinguishable single photons require collection into a single-mode.
In this case, it is the mismatch between the mode profile of the emission impinging upon the fiber facet and the fundamental mode of a single-mode fiber that determines the achievable {\em fiber coupling efficiency} ($\eta_\mathrm{Fib}$).
While the fiber coupling properties of `in-contact' systems such as waveguide to tapered fibers\cite{Patel.2016,Burek.2017} and nanodiamond to tapered fibers\cite{Almokhtar.2014,Vorobyov.2016} have been investigated,  this measure of performance has rarely been considered during the development of nanostructures that rely upon free-space propagation and collection into fiber\cite{Schneider.2018,Rickert.2019}.

This work focuses on exploring the fiber-coupling efficiency of an inverted nanocone (INC) nanostructure, due to promising enhancements of the fiber coupling efficiency offered in this geometry.
This particular nanostructure design has several appealing advantages compared to other designs.
The contributing factors can be conceptually understood in terms of reflection and waveguiding, as illustrated in figure~\ref{fig:INCdesign}(a).
In this context, reflectivity captures the degree of directivity of the emission towards the collection optics ($+z$ direction) instead of radiating into $4\pi$ thus enabling free-space collection efficiencies exceeding 50\%.
The complementary effect of waveguiding can be used to confine emission into a smaller solid angle thereby reducing the diversion of the emitted light.
Furthermore, guiding can also be exploited in order to manipulate the mode shape, which enables higher fiber coupling efficiencies through engineering emission profiles similar to Gaussian beams instead of a dipole profile.
As an example of this, a `trumpet' structure\cite{Munsch.2013,Stepanov.2015}  in semiconductor-quantum dot systems, which has a similar form to the INC but larger heights (greater than 10 $\mathrm{\upmu m}$), has been shown to have a very strong mode-shaping property to obtain such beams with an $\eta_\mathrm{FS}$ of few ten percent without utilizing mirrors below the sample.
For the diamond INC, the reflectivity results from total- and partial internal reflection at the diamond-air interface at the angled walls of the inverted nanocone.
Waveguiding, similarly to nanopillars, arises since the INC can act as a mode-shaping device supporting quasi-normal modes to which the quantum emitter emission can couple.

In this letter the performance of INC structures is investigated by optimizing the design parameters for three alternative metrics for collection performance: {\em fiber coupling}, {\em free-space collection}, and {\em rate enhancement}.
These three different metrics offer different designs with 66\% fiber coupling efficiency, 83\% free-space collection, and 80\% fiber coupling of the emission scaled to the power emitted by a dipole emitter in bulk.\\

Finite element method simulations are used to model fiber coupling as it is well suited for simulating accurately point emitters in complex shaped geometries. 
These evaluations are computed using the nanophotonics simulation package JCMsuite\cite{Burger.2008}.
The fiber coupling simulations consist of computing the emission that scatters from the INC, propagating this through an imaging system, and calculating its overlap with the HE$_{11}$ fiber mode\cite{Schneider.2018}.
The presented results are given with a numerical precision well below 0.01 to ensure numerical stability during the optimization procedures.
Details regarding the simulations, including convergence and computational times, are presented in Supplementary 1.

The INC geometry, as illustrated in figure~\ref{fig:INCdesign}(b), is parametrized in terms of full height ($h$) between the top facet of the inverted nanocone and the sample surface, the top radius ($r_\mathrm{t}$) of the emitting facet of the INC, and the bottom radius ($r_\mathrm{b}$) of the circle connecting the INC to the sample.
The color center is modeled as a point dipole centered on the INC axis ($z$) and aligned in the $x-z$ plane, at a depth of $d_z$ from the top facet.
In subsequent analyses, a lateral displacement, $d_{x}$, of the dipole within the $x-y$ plane is also considered.
An emission wavelength of 619~nm is selected, corresponding to the zero-phonon line (ZPL) transition wavelength of tin-vacancy (SnV) color centers\cite{Iwasaki.2017} due to their high emission into the ZPL ($\sim0.57$ at 10 K)\cite{Gorlitz.2020} in combination with their possible favourable spin coherence properties at temperatures achieved by a standard helium cryostat ($T^*_2=540\pm40\,\mathrm{ns}$ at 2.9 K)\cite{Trusheim.2020}. 
The chosen fiber parameters are based on a 630HP Thorlabs single-mode fiber, and is modeled with a $3.5\,\upmu \mathrm{m}$ diameter fiber core with $n_\mathrm{core}=1.4632$ and an infinitely large cladding with $n_\mathrm{clad}=1.4574$ at 619 nm.

In all simulations, an optical imaging system with a numerical aperture (NA) of 0.9 is used to collect emission from the nanostructure.
The fraction of emission coupled into this solid angle determines $\eta_\mathrm{FS}$.
For fiber-coupling efficiency, an additional lens-system is implemented for further mode matching to the fiber mode. 
A schematic drawing of the collection schemes is shown in figure~\ref{fig:INCdesign}(c).
A comparative discussion on the efficiency of a direct coupling scheme without the imaging system is provided in Supplementary 4.a.

For determining the optimal parameter values for maximal efficiencies, the Bayesian optimizer of JCMsuite is employed.
The method enables a very efficient global search in a high dimensional parameter space with many local minima \cite{Schneider.2019}.
After performing an optimization (labeled $\Lambda$), the set of parameters obtained are referred to with an identification label INC-$\Lambda$.\\

We first consider the performance of an ``ideal'' INC that is optimized for fiber coupling (INC-Fib) by varying over the design parameters in figure~\ref{fig:INCdesign}(b).
For maximal coupling to the INC modes, and thus an optimized $\eta_\mathrm{Fib}$, the dipole orientation is fixed parallel to the sample surface, aligned along the $x$ axis. 
The optimization was performed over $h$, $d_z$, and $r_\mathrm{t}$ in order to maximize $\eta_\mathrm{Fib}$, where $r_\mathrm{b}=1\,\mathrm{nm}$ was set as a constraint.
This study obtained $\eta_\mathrm{Fib}=0.66$ with design parameters $h=635\,\mathrm{nm}$, $r_\mathrm{t}=391\,\mathrm{nm}$, and $d_z=507\,\mathrm{nm}$, resulting in a side-wall angle of $58^\circ$ between the sample surface and the INC.
Further details of the optimization conditions are provided in Supplementary 2.

As discussed, however, free-space coupling is a more commonplace figure of merit.
For comparison with prior work, the optimization procedure is therefore repeated to maximize $\eta_\mathrm{FS}$ instead, yielding design parameters for an INC that is optimized for free-space collection (INC-FS).
In this case $\eta_\mathrm{FS}=0.83$ with $h=511\,\mathrm{nm}$, $r_\mathrm{t}=848\,\mathrm{nm}$, and $d_z=82\,\mathrm{nm}$ resulting in a side-wall angle of $31^\circ$.

While the fiber coupling properties of other diamond nanostructures have not been investigated, the free-space efficiency, $\eta_\mathrm{FS}=0.83$ of the INC-FS demonstrates significant improvement to simulations showing $\eta_\mathrm{FS}=0.298$ for a microlens\cite{Hadden.2010} (NA:0.9, averaged over 600-800 nm) and $\eta_\mathrm{FS}=0.53$ for a nanopillar\cite{Hausmann.2010} (NA:0.9, at 637 nm).
Furthermore, it slightly outperforms $\eta_\mathrm{FS}=0.752$ of an optimized bullseye grating\cite{Zheng.2017} (NA:0.95, at 637 nm) and $\eta_\mathrm{FS}=0.77$ of a parabolic reflector\cite{Wan.2018} (NA:1.3, averaged over 600-800 nm).

\begin{figure}[t]
\includegraphics[width=\linewidth]{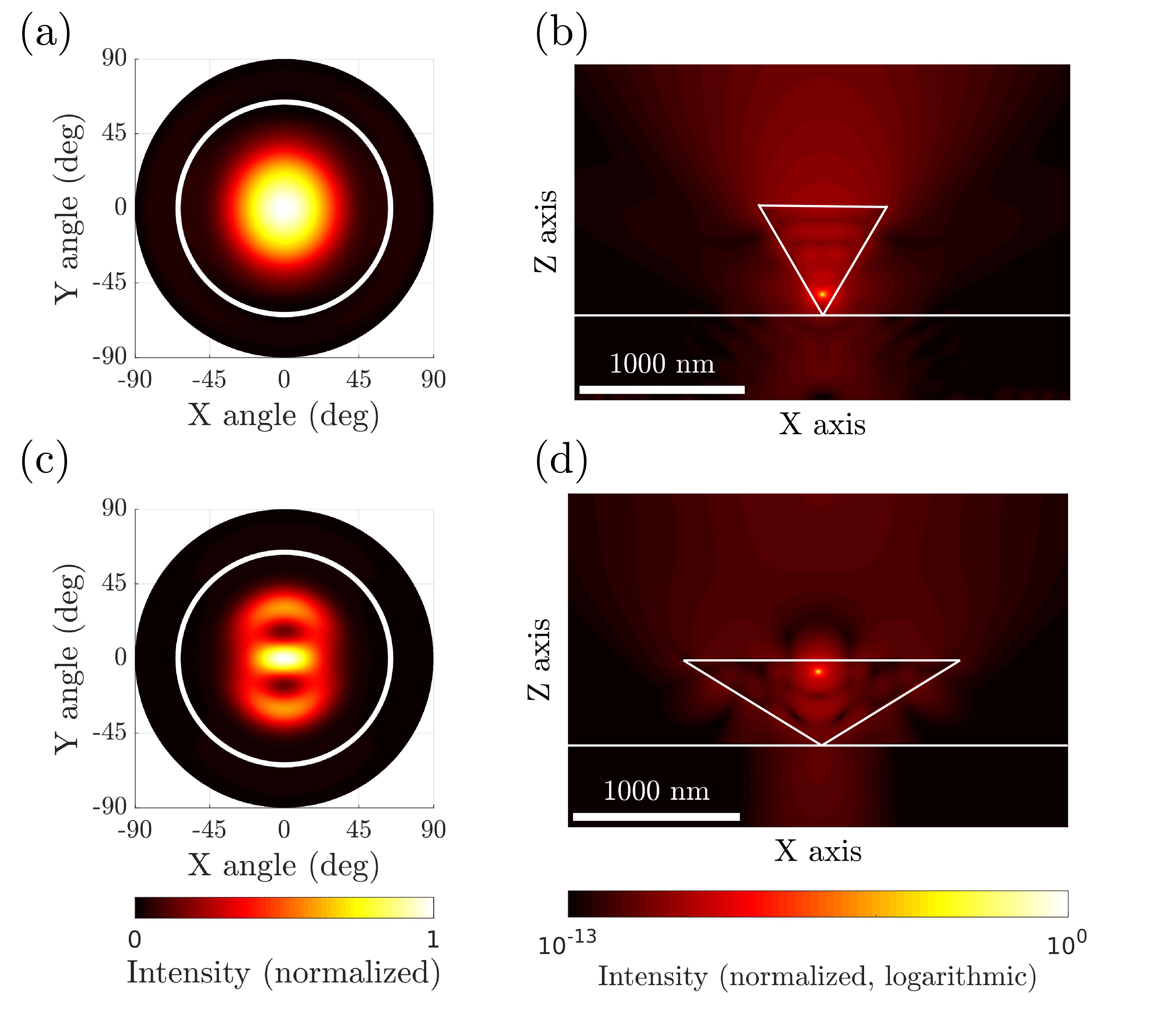}
\caption{
Far-field emission profiles and field distributions along the computational domain for a dipole radiating at 619 nm.
(a) Far-field emission profile and (b) field distribution resulting from an INC optimized for fiber coupling (INC-Fib).
(c, d) Corresponding field profiles for an INC optimized for free-space coupling (INC-FS).
The circle in the far-field profiles (a, c) indicates the solid angle from which emission can be collected, and is used to determine the free-space coupling efficiency.
The outline in the field distribution sections (b, d) highlights the outer dimensions of the INC and sample substrate.
}
\label{fig:fields}
\end{figure}

From the definitions of the alternative efficiency measures, both $\eta_\mathrm{FS}$ and $\eta_\mathrm{Fib}$ necessitate emission into the solid angle determined by the collection optics.
Fiber coupling, however, places additional requirements on the mode quality of emission.
Consequently, structures optimized for fiber coupling efficiency therefore also enable high free-space collection efficiency, whereas designs considering only free-space collection will not necessarily couple well to a fiber mode (see Supplementary 4.b).
In the far-field, it is readily seen that INC-Fib produces a near uniform Gaussian beam profile, figure~\ref{fig:fields}(a), whereas INC-FS exhibits a multimode profile with significant sidelobes, figure~\ref{fig:fields}(c), due to the larger top radius.

Since $\eta_\mathrm{FS}$ is insensitive to the emitted mode profile, the convergence to a large top radius of the INC-FS is driven by increasing the directivity of the emission in $+z$, through increasing the opening angle of the cone and hence, the fraction of emission fulfilling the total internal reflection condition (seen in figure~\ref{fig:fields}(d) showing the INC cross-section and the local field distribution).
Conversely, INC-Fib tends towards a larger $d_z$ and reduced $r_\mathrm{t}$ which have the effect of increasing the waveguiding effect of the INC, which together result in a near single-mode emission profile figure~\ref{fig:fields}(b).\\

Having identified the essential elements for the design of INC structures for optimized collection efficiencies, the feasibility of fabricating devices in practice must also be considered.
Fabrication processes introduce uncertainties in the nanostructure shape and dimensions, which in turn impact the final performance of a device.
In the following analysis, a fabrication precision of $\sim10\,\mathrm{nm}$ is assumed to be achievable.

To provide an estimate of tolerances during fabrication, a {\em fabrication insensitivity score} $\mathcal{S}(\zeta)$ is obtained from the variation of a chosen performance metric $\zeta$, as each of the geometrical parameters of a given INC-$\Lambda$ is individually swept about the target value, over a range reflecting physically achievable precision, while maintaining other dimensions constant.
Further information regarding how these metrics are obtained are detailed in Supplementary 3.

Evaluation against this metric yields similarly high values for both structures considered, with $\mathcal{S}(\eta_\mathrm{Fib})=0.92$ for INC-Fib, and $\mathcal{S}(\eta_\mathrm{FS})=0.94$ for INC-FS.
This suggests that the INC structure may offer a repeatable -- and therefore scalable -- design for enhancing the achievable collection efficiencies from quantum emitters.\\

\begin{figure}[t]
\includegraphics[width=\linewidth]{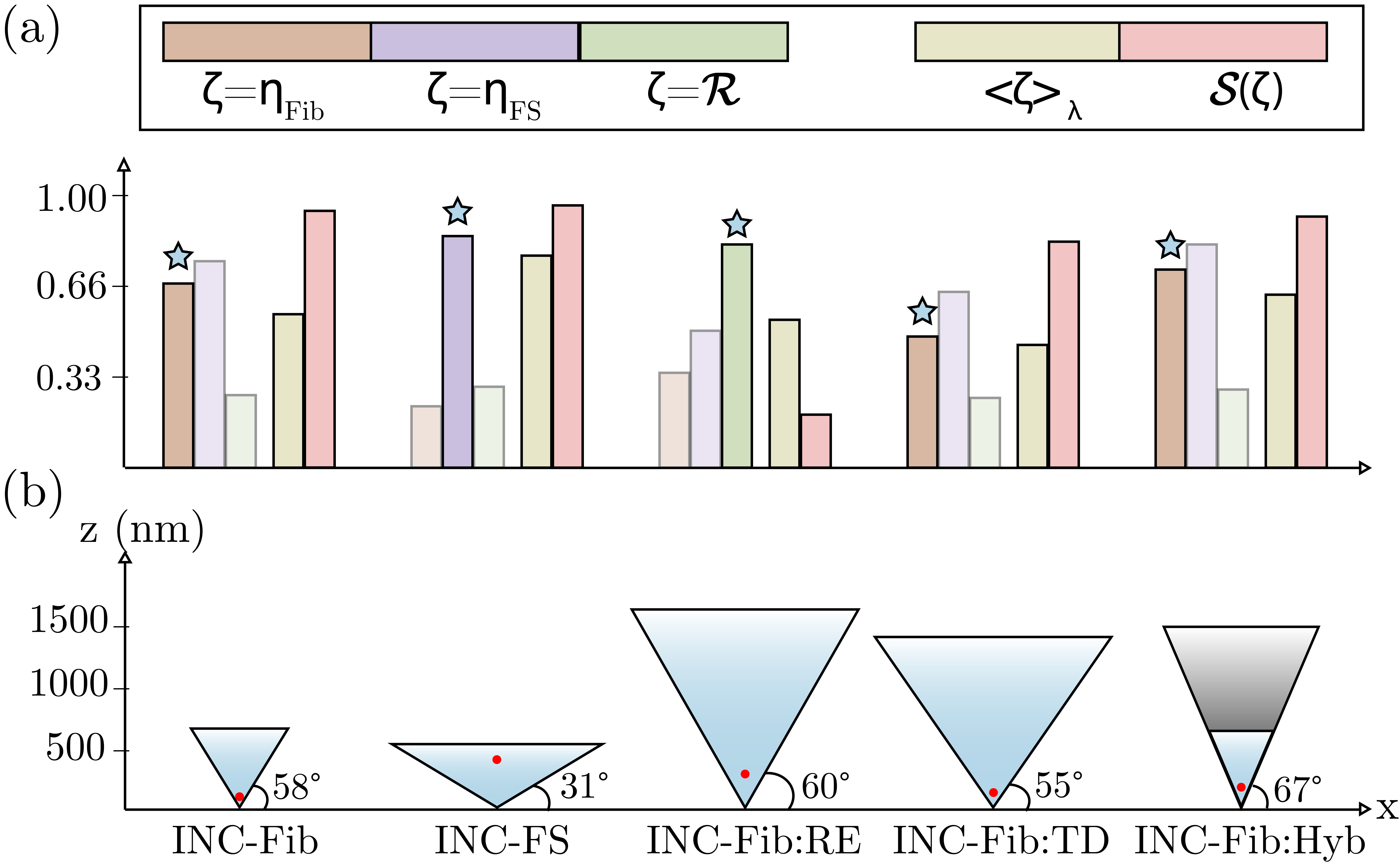}
\caption{
Comparison of the resulting performance and geometries of INC optimized for different figures of merit.
A full tabulation of results can be found in table~\ref{tab:INC-comparison}.
(a) Bar chart representing the performances of the five optimized INCs. The INC parameters are optimized using the figure of merit $\zeta$ indicated by the star. Further performance metrics $\left\langle\eta_\mathrm{Fib}\right\rangle_\lambda$ (light yellow bar) and $\mathcal{S}(\zeta)$ (pink bar) are evaluated using $\zeta$.
(b) Geometries of the five INC plotted up to scale with inscribed side-wall angles.
}
\label{fig:outputnumbers}
\end{figure}

Depending on the application, the total collectible emission rate rather than purely the fraction of light coupling into a single-mode fiber may be a more significant measure of performance.
The local density of states is modified by the INC and determines the absolute fluorescence rates of the color centers through the Purcell effect.
In comparison to bulk diamond, the effective refractive index of the INC is in general reduced, since the mode is distributed between diamond and air; hence, unless a cavity effect is achieved in an INC, its fluorescence emission rate is suppressed with respect to emitters in bulk.
The ratio of emission inside the nanostructure and the bulk is expressed by the Purcell factor ($F$).
The INC-Fib, optimized for fiber coupling, has a $F = 0.39$, whereas the INC-FS achieves a weak cavity effect with $F = 1.32$.

Analogous to brightness enhancement\cite{Abudayyeh.2017} for free space coupling, an alternative figure of merit may be employed to quantify the {\em fiber-coupled rate enhancement}, i.e. the fraction of fiber coupled light under normalization against a dipole emitter in bulk diamond.  The rate enhancement therefore provides a means to optimize the total usable count rate.
In this case, we define $\mathcal{R}=F\eta_\mathrm{Fib}$ to measure the rate enhancement in conjunction with fiber coupling.

Optimizing the INC parameters for brightness with fiber coupling (INC-Fib:RE) yields $\mathcal{R}=0.80$, with $\eta_\mathrm{Fib}=0.34$, and $F=2.34$, which constitutes a factor of $\sim3$ improvement in rate compared with the INC-Fib ($\mathcal{R}=0.26$); however, since the rate increase originates from a cavity enhancement, the performance becomes more sensitive to any variations in the structure geometry -- thus introducing challenges in fabrication -- seen in $\mathcal{S}(\mathcal{R})=0.19$ for INC-Fib:RE, as compared to $\sim0.9$ for INC-Fib and INC-FS.
\\

A further important consideration is the impact of dipole orientation within the nanostructure.
To obtain a horizontally oriented dipole for SnV color centers, a (110) terminated surface crystal could be used, in which two of the four SnV dipole orientations within the lattice ($\left[1\overline1\overline1\right]$, $\left[\overline11\overline1\right]$) are parallel to the diamond surface.
However, many commercial diamond samples have a (100) terminated surface, for which all 4 SnV dipole orientations, $\left\langle111\right\rangle$,  are equivalent, and result in an angle of 35.3$^\circ$ between the dipole axis and the (100) plane. 

Optimizing for $\eta_\mathrm{Fib}$ with the tilted dipole (TD) orientation ($\left\langle111\right\rangle$ dipole in $(100)$ crystal) yields the parameters for INC-Fib:TD, for which $\eta_\mathrm{Fib}=0.47$. 
The approximately 20\% reduction of efficiency with respect to the $\eta_\mathrm{Fib}=0.66$ for INC-Fib arises from the reduced symmetry, which therefore impacts the far-field mode quality.
The optimization attempts to compensate for this and converges to approximately twice the height of INC-Fib in order to increase the wave-guiding property of the structure, whilst retaining similar proportions in terms of the INC opening angle and relative depth of the emitter.
Conversely, if the dipole in INC-Fib:TD is horizontal instead, then the fiber coupling increases to $\eta_\mathrm{Fib}=0.60$, near the optimized case.

The fabrication insensitivity without changing $d_x$ remains reasonably robust with $\mathcal{S}(\eta_\mathrm{Fib}) = 0.81$, but is more sensitive to $x$ position due to the off-axis dipole.
An analysis of this effect and further comparative discussions of the optimized INCs are provided in Supplementary 4.d-e.
\\

\begin{figure}
\includegraphics[width=\linewidth]{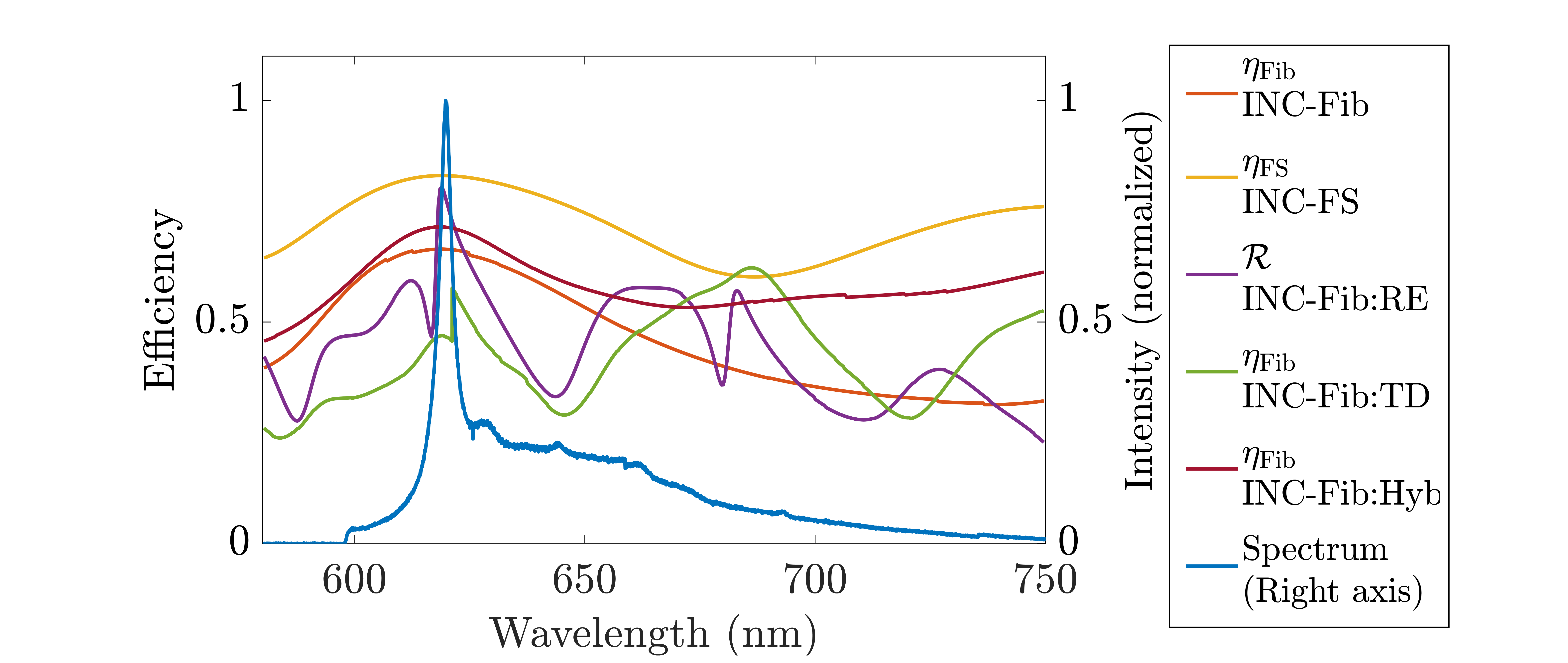}
\caption{
Efficiencies achieved by the optimized inverted nanocones at different wavelengths and the tin-vacancy color center fluorescence spectrum measured at room temperature under 520 nm non-resonant excitation.
}
\label{fig:iin_bbe}
\end{figure}

While many quantum information processing applications require indistinguishable photons collected from the zero-phonon line, other applications may make use of the full emission spectrum.
We therefore define a measure of the broadband performance for metric $\zeta$ of a given structure by averaging over the spectral intensity, $I(\lambda)$, of the emitter:
$\left\langle\zeta\right\rangle_\lambda=\int\!\mathrm{d}\lambda\,\zeta(\lambda)I(\lambda),$
where $I(\lambda)$ is taken from a measurement of an SnV fluorescence spectrum measured at room temperature, in the region 580-750~nm (see  Supplementary 5 for experimental details).
Using this measure, INC-Fib achieves a broadband fiber coupling efficiency $\left\langle\eta_\mathrm{Fib}\right\rangle_\lambda=0.55$.

Figure \ref{fig:iin_bbe} shows $I(\lambda)$, and $\zeta(\lambda)$ for the different INC designs.
In most cases, the performance is only weakly modulated as a function of the wavelength (see Supplementary 4.f); however, a strong modulation is observed for  INC-Fib:RE.
Since the Purcell enhancement necessitates coupling to quasi-normal modes (QNMs) of the structure, the stronger oscillatory behavior can be attributed to the presence of QNMs with narrow bandwidths. 
Similar modulations are exhibited for the INC-Fib:TD, which has a comparable geometry to INC-Fib:RE, and may support similar modes, and thus has a stronger waveguiding effect in contrast to the shorter INCs.
Furthermore, the asymmetry of source within the INC could allow for the possibility of weakly coupling with more QNMs, and correspondingly has less trivial wavelength dependence of the final mode-mismatch with the single-mode fiber.
 \\

As a final remark, the waveguiding property of an optical structure is not only a consequence of the geometry but also the material refractive index.
A supplementary approach to engineering the guiding properties is to consider a hybrid structure incorporating multiple materials.
As a simple example of this, a multi-material INC is considered, in which the top section of the INC above the emitter is substituted for a higher $n$ material, resulting in stronger confinement of the light and thereby directing emission towards the upper half.
Here we consider a layer of a conceptual material with $n=3.65$, deposited onto a bulk diamond sample, and an INC fabricated out of both materials together.
In practice, AlGaAs may achieve this value\cite{Aspnes.1986} while GaP has a similar refractive index as high as $n=3.33$\cite{Aspnes.1983} and both materials could in principle be used for fabrication.

To provide an indicative analysis, an analogous optimization to INC-Fib was performed with the additional free parameter of the high-$n$ coating thickness.
Optimisation (INC-Fib:Hyb) achieves both an improvement in fiber coupling to the pure diamond INC-Fib with $\eta_\mathrm{Fib}=0.71$, while also reaching a comparable free-space efficiency to the INC-FS with $\eta_\mathrm{FS}=0.80$.
Furthermore, the fabrication insensitivity of fiber coupling also remains high with $\mathcal{S}(\eta_\mathrm{Fib})=0.90$.
While such a hybrid material system has not been fabricated and it is therefore difficult to predict its feasibility, this design offers an effective and relatively simple means for further enhancing the collection efficiency from color centers in diamond compared to pure diamond designs.
\\

In conclusion, the fiber coupling properties of an optical nanostructure, the INC, is investigated. 
This detailed numerical investigation focuses particularly on fiber-coupling as an important – and often neglected – figure of merit that is requisite for providing indistinguishable single modes for the generation of indistinguishable photons, as well as a means for coupling multiple quantum photonic devices in a modular and scalable way.
Although defects in diamond have been considered here, the methods presented are more generally applicable to other quantum emitters and host materials, which require free-space collection in their operation.

INC structures providing typically 60\% fiber coupling efficiency (70\% free-space coupling efficiency) were found, and indicate promising tolerance to fabrication errors.
A summary of results and comparisons is tabulated in table~\ref{tab:INC-comparison}.

\begin{table}[]
\setlength{\tabcolsep}{1ex}
\begin{tabular}{
l
@{\hspace{2\tabcolsep}} *{4}{r}
@{\hspace{2\tabcolsep}} *{4}{r}
}
\toprule
\multirow{2}{*}{INC-$\Lambda$}
& \multicolumn{1}{c}{$\eta_\mathrm{fib}$}
& \multicolumn{1}{c}{$\eta_\mathrm{fs}$}
& \multicolumn{1}{c}{$\mathcal{R}$}
& \multicolumn{1}{c}{$\left\langle\zeta\right\rangle_\lambda$}
& \multicolumn{1}{c}{$h$}
& \multicolumn{1}{c}{$r_\mathrm{t}$}
& \multicolumn{1}{c}{$d_z$}
& \multicolumn{1}{c}{$\mathcal{S}(\zeta)$}\\
& & & &
& \multicolumn{1}{c}{\footnotesize{(nm)}} & \multicolumn{1}{c}{\footnotesize{(nm)}} & \multicolumn{1}{c}{\footnotesize{(nm)}}
& \\
\midrule
Fib\footnotemark[1]
& \hlem{0.66} & 0.74 & 0.26 & 0.55
& 635 & 391 & 507
& 0.92 \\
FS\footnotemark[2]
& 0.22 & \hlem{0.83} & 0.29 & 0.76
& 511 & 848 & 82
& 0.94
\\
Fib:RE\footnotemark[3]
& 0.34 & 0.49 & \hlem{0.80} & 0.53
& 1 599 & 914 & 1 286
& 0.19 \\
Fib:TD\footnotemark[1]
& \hlem{0.47} & 0.63 & 0.25 & 0.44
& 1 375 & 953 & 1 261
& 0.81 \\
Fib:Hyb\footnotemark[1]\footnotemark[4]
& \hlem{0.71} & 0.80 & 0.28 & 0.62
& 1 462 & 624 & 1 295
& 0.90 \\
\bottomrule
\end{tabular}
\footnotetext[1]{Evaluated with $\zeta=\eta_\mathrm{Fib}$.}
\footnotetext[2]{Evaluated with $\zeta=\eta_\mathrm{FS}$.}
\footnotetext[2]{Evaluated with $\zeta=\mathcal{R}$.}
\footnotetext[4]{Total INC dimensions, where $h_\mathrm{diamond}=855\,\mathrm{nm}$; $h_\mathrm{coating}=607\,\mathrm{nm}$.}
\caption{
Performance metrics and corresponding dimensions for INCs optimized according to different figures of merit \hlem{(purple text)} for a dipole radiating at 619 nm.
After optimization, performance against alternative metrics is evaluated and given.
}
\label{tab:INC-comparison}
\end{table}

These results highlight that while free-space coupling only considers directivity, fiber coupling additionally requires improved mode-quality of emission, and is a more practical figure of merit in quantum technology applications.
Optimization of inverted nanocones for fiber coupling generally finds wide side-wall angles between $\sim55-70^\circ$, with the emitter near the base of the structure at depths of $\sim80-90\%$ of the full structure height.
The initial indicative analyses on its tolerance to fabrication errors show that within reasonable fabricable precision of 10 nm the performance remains robust.
    The INC, therefore, presents an appealing design for the enhancement of photon collection from color centers in diamond, due to its favorable geometry which simultaneously enables high directivity, controlled divergence, and single-mode emission, that is provided in a package promising fabrication simplicity, and is readily integrable into cryogenic setups.

\section*{SUPPLEMENTARY MATERIAL}

See supplementary material for details regarding the simulation steps, optimizations, fabrication insensitivity analysis, further discussions and experimental acquisition of the spectrum.

\begin{acknowledgements}
The authors would like to thank Tommaso Pregnolato, Gregor Pieplow, and Julian Bopp for insightful discussions and technical assistance.
This research was supported through the Federal Ministry of Education and Research of Germany (BMBF, project DiNOQuant, 13N14921) and the European Research Council (ERC Starting Grant “QUREP”). 
P.I.S., M.H. and S.B. acknowledge support by BMBF Forschungscampus MODAL, project number 05M20ZBM.
JHDM gratefully acknowledges support from the Alexander von Humboldt Foundation.

\end{acknowledgements}

\section*{DATA AVAILABILITY}
The data that support the findings of this study are available from the corresponding author upon reasonable request.

\end{document}